\documentclass[final,journal,letterpaper]{IEEEtran}

 \usepackage[hidelinks]{hyperref}
 \usepackage{varioref} 
 \usepackage{wrapfig} 
 \usepackage{threeparttable} 
 \usepackage{dcolumn} 
  \newcolumntype{d}{D{.}{.}{-1}}
 \usepackage{nomencl} 
  \makeglossary
 \usepackage{subfigmat} 
 \usepackage{fancyvrb} 
  \fvset{fontsize=\footnotesize,xleftmargin=2em}
 \usepackage{lettrine} 

 \usepackage{epsfig}
 \usepackage{epstopdf}
 \usepackage{subfigure}
 \usepackage{latexsym}
 \usepackage{amsmath}
 \usepackage{amsfonts}
 \usepackage{amssymb}
 \usepackage{mathrsfs}
 \usepackage{cases}
 \usepackage{array}
 \usepackage{color}
 \usepackage{soul}
 \usepackage{verbatim}
 \setulcolor{red}
 \sethlcolor{yellow}

 \usepackage{url}
 \usepackage[noadjust]{cite}
 \usepackage{graphicx}
 \usepackage{psfrag}
\usepackage{soul}
\usepackage{tikz}
 \usepackage{color}
   \usepackage{multirow}
 \usepackage{ragged2e}
 \usepackage{graphicx}  
 \usepackage{breqn}
 \usepackage{algorithm} 
\usepackage{algpseudocode} 
 
\newcommand\copyrighttext{%
  \footnotesize This work has been submitted to the IEEE for possible publication. Copyright may be transferred without notice, after which this version may no longer be accessible.}
\newcommand\copyrightnotice{%
\begin{tikzpicture}[remember picture,overlay]
\node[anchor=south,yshift=10pt] at (current page.south) {\fbox{\parbox{\dimexpr\textwidth-\fboxsep-\fboxrule\relax}{\copyrighttext}}};
\end{tikzpicture}%
}

\begin{document}

\title{Data-Driven Linear Koopman Embedding for Networked Systems: Model-Predictive Grid Control}

\author{Ramij R. Hossain~\IEEEmembership{Student Member,~IEEE}, Rahmat Adesunkanmi, 
        Ratnesh Kumar,~\IEEEmembership{Fellow,~IEEE}
\thanks{The work was supported in part by the National Science Foundation under the grants, NSF-CSSI-2004766 and NSF-PFI-2141084.
\newline \indent R. R. Hossain, R. Adesunkanmi and R. Kumar are with the Department of Electrical and
Computer Engineering, Iowa State University, Ames, IA 50011, USA (e-mail:
rhossain@iastate.edu, rahma@iastate.edu, rkumar@iastate.edu).}
}

\maketitle
\copyrightnotice
\begin{abstract}
This paper presents a data-learned linear Koopman embedding of nonlinear networked dynamics and uses it to enable real-time model predictive emergency voltage control in a power network. The approach involves a novel data-driven ``basis-dictionary free" lifting of the system dynamics into a higher dimensional linear space over which an MPC (model predictive control) is exercised, making it both scalable and rapid for practical real-time implementation. A {\em Koopman-inspired deep neural network} (KDNN) encoder-decoder architecture for the linear embedding of the underlying dynamics under distributed controls is presented, in which the end-to-end components of the KDNN comprising of a triple of transforms is learned from the system trajectory data in one go: A Neural Network (NN)-based lifting to a higher dimension, a linear dynamics within that higher dimension, and an NN-based projection to the original space. This data-learned approach relieves the burden of the ad-hoc selection of the nonlinear basis functions (e.g., polynomial or radial) used in conventional approaches for lifting to higher dimensional linear space. We validate the efficacy and robustness of the approach via application to the standard IEEE 39-bus system.

\end{abstract}

\section{Introduction} \label{1}
\IEEEPARstart{C}{omplex} nonlinear networked systems abound us, presenting new challenges for their control design in terms of scalability and rapid online computation for real-time execution. These include the modern power grids with massive integration of renewable generations, dynamic loads, and inverter-based resources that necessitate reliable, fast, and real-time control to mitigate instability following any disturbances during the system operations. Another consequence of systems' complex, nonlinear, high-dimensional dynamics is that model-based control designs have time time-consuming and not amenable to real-time execution. To this end, the emerging data-driven control approaches offer opportunities as a potential alternative to the conventional model-based control designs in complex networked cyber-physical systems (CPSs) \cite{radanliev2021artificial,de2019formulas}: A data-driven approach replaces the need for a comprehensive system model and makes the control computation fast and real-time, trading a small modeling accuracy for a substantial gain in computation speed. 

The advent of deep learning technologies combined with reinforcement learning (RL) has started to be explored in various domains, including in power system controls \cite{zhang2019deep}. In deep reinforcement learning (DRL) applications to power systems, \cite{chen2020model} utilized a deep deterministic policy gradient (DDPG)-based algorithm for the frequency regulation problem. A deep Q-network (DQN)-based algorithm for load shedding-based voltage control is reported in \cite{huang}. However, the training of DRL-based controllers has certain challenges when applying to real-world power systems \cite{chen2021reinforcement}: (a) the exploration/training becomes un-scalable with the increase of state and action space, and (b) DRL-based control lacks proven robust performance and safety guarantee, opening the potential for catastrophic failures in real-world applications \cite{chen2021reinforcement}. 

An appealing alternative to learned controls is to employ learned models and perform model-based control designs. But while the design of model-based control, including Model Predictive Control (MPC), is scalable for linear dynamics, and while MPC has shown promise in power system optimal regulation \cite{cutsem_review,jin2009model,hiskens2006voltage}, it is well accepted that designing MPC for a real-time power system is challenging, even when incorporating different approximation strategies, for instance, trajectory sensitivity-based methods \cite{hiskens2000trajectory,jin2009model}.

Motivated by this, we present a method of linear lifting of nonlinear system dynamics and leverage it to design MPC over the lifted linear dynamics, making the entire setup scalable and real-time, overcoming the primary drawback of the otherwise promising MPC for complex CPSs. Our method is inspired by linear Koopman lifting of nonlinear dynamics, wherein the underlying distributively controlled dynamics of a networked system is learned in the form of the proposed encode-decoder neural-network architecture embedding an interim linear stage, termed KDNN (Koopman-inspired deep neural network). An advantage of the approach is that it avoids the need to know the basis functions to lift the given nonlinear dynamics to a linear space or to project it back, rather learns those using the trajectory data of the underlying system. The MPC computation is then directly performed on the linear embedding, which makes it scalable and real-time. The approach presented applies to any networked system as shown by the case for power grids analyzed in this work. The unique contributions are data-learned KDNN representation for a nonlinear networked system involving a linear embedding that auto-learns the lifting and projecting basis functions, and the resulting speed-up of distributed control computation---a 35-fold reduction in comtrol computation time is achieved for the IEEE 39-bus power network examined in the paper, making MPC real-time for a first time for power networks.

\subsection{Related Literature}
Recent advances in data-driven methods have made Koopman operators a leading candidate for the linear representation of nonlinear systems \cite{koopman1931hamiltonian}. According to Koopman theory, the nonlinear dynamical system is lifted to a higher-dimensional space using nonlinear basis functions, where the evolution of lifted states is governed by a linear operator, known as the Koopman operator, and whose projection to the original space provides an approximation to the original trajectory of the nonlinear system \cite{williams2015data,korda2018linear}. While obtaining a linear embedding is promising, lifting nonlinear dynamics to a higher dimensional linear space is nontrivial since an appropriate set of nonlinear basis functions must be identified first: Standard approaches utilize some predefined dictionary of basis functions, for instance, radial or polynomial basis functions for such lifting, that need be the most effective. 

The conventional Koopman-based identification and control design methods have also been explored in power systems applications. \cite{korda2018power} solves an MPC-based transient stabilization problem utilizing classical swing dynamics of power systems \cite{kundur1994power}. A Koopman based model predictive power system stabilizer (PSS) is designed in \cite{han2021koopman}. \cite{han2022model} integrates Koopman-based system identification and controller design for oscillation damping. Koopman-based controller for decentralized frequency control problem with the time-delayed embedding of measurements is studied in \cite{li2021determination}. In \cite{sharma2021data}, data-driven characterization of power system dynamics is investigated using Koopman operator theory. 

The Koopman-based embedding in power systems is mostly based on simplified models of power system dynamics so that a predefined dictionary of basis functions (e.g., equation (2) of \cite{korda2018power}) can be useful. In general, however, power systems obey differential-algebraic equations (DAEs) based evolution, and it is necessary to include nonlinear dynamics comprising higher-order generator models (instead of the classical model), exciter models, load models, etc., particularly for voltage stability issues. To this end, building an appropriate dictionary of basis functions for a complex nonlinear DAE model is challenging, and it restricts the application of Koopman-based methods in power system control design. This issue has also been observed in other complex nonlinear applications. 

To alleviate this problem, recent researchers have leveraged advances in artificial intelligence (AI), particularly NN, to learn the basis functions required to lift the nonlinear dynamics and also learn the linear Koopman embedding \cite{enoch,lusch2018deep,robotics1,robotics2,auto1, vehicle,shi2022deep}. The advantage of the learning-based lifting of nonlinear dynamics has also been recognized in power systems: \cite{powersys} utilized this approach for transient stability based on a simplified model of the swing dynamics, similar to the simplified model used in \cite{korda2018power}. Instead of using predefined basis functions, \cite{powersys} utilizes neural networks to learn the basis functions for lifting operations. The Koopman operators are identified by solving a least square estimation problem in conjunction with the iterative learning process of the basis functions. However such iterative learning runs the risk of numerical issues between the iterations of learning the neural basis functions and the least square estimation of Koopman operators.

\subsection{Our Approach}
In this work, we develop a data-learned linear Koopman embedding framework and employ it for an MPC-based control. Recognizing that through the system DAEs, the bus voltages and reactive compensation are related by an implicit nonlinear function (in contrast to the explicit relationship among the voltages, the rotor angles, and the mechanical power inputs of the DAEs). We present a Koopman-inspired deep neural network architecture, termed KDNN, for a linear embedding of this implicit nonlinear voltage dynamics, by drawing data from the system evolution based on its full differential-algebraic dynamics (without any approximation). A distinguishing feature of our work is that it utilizes an encoder-decoder structure comprising  (a) NN-based auto-learned basis functions for lifting, (b) auto-learned linear Koopman operators associated with lifted states, and control inputs, and (b) auto-learned projection function used for mapping down to the original state-space. Also, this encoder-decoder-based neural architecture is trained in an end-to-end manner learning the unknown basis function, the linear Koopman operators, and the projection function together in one go. Such end-to-end learning eliminates the aforementioned risks of numerical issues associated with iterative learning of neural basis functions vs. the least square estimation of Koopman operators, as is the case in \cite{powersys}. Some prior works support our choice of encoder-decoder KDNN structure \cite{lusch2018deep, vehicle} by way of applications in domains such as modeling vehicle dynamics or general nonlinear systems. 

The validity of the proposed KDNN-based MPC scheme is established by implementing an emergency voltage recovery scheme following a severe fault in the IEEE 39-bus system. The performance and robustness of the framework are also demonstrated via $\pm 10 \%$ load variations around the nominal model. The same model was also used to learn the KDNN-based linear embedding.

\subsection{Main Contributions}
\begin{enumerate}
    \item A  Koopman-based,  data-driven, ``basis-dictionary free", linear embedding of a nonlinear networked dynamics, that comprises a deep learning-based end-to-end encoder-decoder architecture, termed KDNN, for lifting the controlled nonlinear dynamics to a higher dimensional linear space. 
    \item An end-to-end KDNN training methodology that only utilizes the trajectory data collected from the controlled nonlinear networked system dynamics under a variety of inputs and disturbances, to {\em simultaneously} learn the (i) lifting bases, (ii) linear dynamics of the lifted domain, and (iii) projection mapping to the original space.  
    \item Validation of the proposed approach for both efficacy and robustness by application to IEEE 39-bus power network and $\pm$~10\% load variation, and its comparison with standard Koopman-based approach involving radial or polynomial basis.
\end{enumerate}

\section{Proposed Methodology}\label{2}
We start by considering the nonlinear power system dynamics representation in the DAE form that implicitly models the bus voltage dynamics under the corresponding control. We then provide a brief overview of the Koopman operators and present the proposed data-driven Koopman-inspired DNN architecture (KDNN) for embedding the nonlinear voltage dynamics into a higher dimensional linear space. Finally, we present the model predictive control framework in the lifted linear space to stabilize voltage trajectories following any large disturbance in the system dynamics caused, for example, by a line fault. 

\subsection{Power System Model}
The dynamics of a power system can be captured in the form of a nonlinear DAE as given in (\ref{sysdae}):
 \begin{equation}\label{sysdae}
 \dot{X}=F(X,Y,U);\;\; 0=G(X,Y,U),
       \end{equation}
\noindent where $X:=$ state variables, $Y:=$ algebraic variables, and $U:=$ control inputs. Here we are interested in the voltage stability control, so the relevant dynamics is the response of bus voltages $V \subseteq Y$ to the control input $U$ (reactive VAR compensation in our setting). As noted above, the use of the nonlinear model (\ref{sysdae}) is not scalable for real-time MPC-based voltage stabilization, and instead, we look to learn the relevant part of the dynamics, namely the distributively controlled voltage dynamics, against that data obtained from the DAE model trajectories in the proposed KDNN framework. 

Since the system trajectory data is available in discrete time, we examine the voltage dynamics in discrete time, indexed by integer variable $k\geq 0$, as:
 \begin{equation}\label{voleqn}
 V_{k+1}=\mathcal{T}(V_{k},U_{k}),
 \end{equation}
where $V_{k}:=[V^1_k,\cdots,V^n_k]^T$ is the voltage vector and $U_{k} := [U^1_k,\cdots,U^m_k]^T$ is the control vector at $k^{\text{th}}$  control instant in a $n$-bus power grid with $m$ number of control inputs. In a practical setting, the control implementation occurs at a larger time period $T_c$, compared to the voltage discretization time period, $T_{s}$. We let $H=T_c/T_s$ denote the length of voltage trajectory history between any two control instants. The controls are held constant between the two control instances while the voltage dynamics continue to evolve, forming a time series of length $H$ between any two control instants. Thus, at each control instant $k$, the ``voltage-state" is the time-series of voltage values between the control instants $k-1$ and $k$, that can be captured in a matrix form: 
\begin{align} \label{volhist}
   V^{\text{hist}}_{k} =  \begin{bmatrix}
V^1_{k,0} & \cdots & V^1_{k,H-1}\\
\vdots & \vdots & \vdots\\
V^n_{k,0} & \cdots & V^n_{k,H-1} \end{bmatrix}.
\end{align}
In contrast, the control input $U_k$ only takes a new value at the instant $k$. Henceforth, with a slight abuse of notation, we use $V_k$ to represent $V^{\text{hist}}_{k}$. 

Note that the DAEs (\ref{sysdae}) are an interacting system of simultaneous equations written compactly employing vectors: Its state equations $F$ are for the vector of all the node states, whereas the vector of sum total of all the networking constraints of voltages and powers form the algebraic equations $G$. The underlying voltage dynamics, as a function of controls distributed over the entire network, in vector form is given as (\ref{voleqn}), whose dimension is the number of network nodes. A novelty is that the ``voltage-state" for each node is taken to be a time-series of voltage values between any two control instants (since voltage sampling rate is higher than the control application rate). 

Next we discuss the lifting of the nonlinear mapping $\mathcal{T}(\cdot,\cdot)$ of (\ref{voleqn}) to a higher dimensional linear state space by way of employing the theory of Koopman operators.

\subsection{Koopman Operator and Lifting of Voltage Dynamics}
Let's consider a discrete time nonlinear dynamics:
 \begin{gather}
 x_{k+1}=f(x_{k},u_k) \label{geneqn2}
 \end{gather}
where, $x\in \mathbb{R}^n$, $u\in \mathbb{R}^m$, and $f$ is the vector field. We adopt the lifting mechanism presented in \cite{korda2018linear} which only lifts the state variables $x$ to a higher dimension, leaving the control variables $u$ unlifted. According to koopman operator theory for finite dimensional linear approximation of nonlinear system defined by (\ref{geneqn2}), there exist $N >> n$, real-valued nonlinear basis functions (lifting functions) ${\mathcal{G}}_i:{\mathbb{R}}^n\rightarrow\mathbb{R}$ for $i=1,\cdots,N$ forming ${\mathcal{G}}=[{\mathcal{G}}_1,\cdots,{\mathcal{G}}_N]^\top$, which lifts the original state space to the higher dimensional state-space so that in the lifted space, the system follows the linear dynamics:
 \begin{equation}\label{koopcon}
\mathcal{G}(x_{k+1})\;\;=\;\;\mathcal{A}\; \mathcal{G}(x_{k}) \;+\; \mathcal{B}\; u_{k},
\end{equation}
where $\mathcal{A} \in {\mathbb{R}}^{N\times N}$, and $\mathcal{B} \in {\mathbb{R}}^{N\times m}$ are the Koopman operators associated with the state and control spaces respectively. 

\subsubsection{Standard EDMD for linear system estimation \cite{korda2018linear}}
One way to obtain the Koopman operators for a finite dimensional approximation such as (\ref{koopcon}) is to utilize the method of extended dynamic mode decomposition (EDMD) \cite{williams2015data,korda2018linear}, where a predefined dictionary of basis functions is assumed known. The approach requires collecting $T$ tuples $\{(x_k,u_k,x_{k+1})_{k=1}^T\}$ from each trajectory data of the given nonlinear system under different initial conditions. The collected trajectory data is arranged as follows:
\begin{gather*}
    \mathbf{X} = [x_1,\cdots,x_T], \mathbf{X}^{+} = [x_2,\cdots,x_{T+1}], \mathbf{U} = [u_1,\cdots,u_T],
\end{gather*}
where the $i^{\text{th}}$ elements $\mathbf{X}(i), \mathbf{X}^{+}(i)$, and $\mathbf{U}(i)$ satisfy the dynamics (\ref{geneqn2}), namely, $\mathbf{X}^{+}(i) = f(\mathbf{X}(i),\mathbf{U}(i))$.
For a given dictionary of mappings,
\begin{gather*}\label{std1}
        \bar{{\mathcal{G}}}(x) = \begin{bmatrix}
    \bar{{\mathcal{G}}}_1(x)\\
\vdots \\
\bar{{\mathcal{G}}}_N(x)
    \end{bmatrix},
\end{gather*}
the lifted space representations are:
\begin{gather*}\label{std2}
   \mathbf{X}_{\text{lift}} = \begin{bmatrix}\bar{{\mathcal{G}}}(x_1)&\cdots&\bar{{\mathcal{G}}}(x_T)\end{bmatrix};\\
      \mathbf{X}^{+}_{\text{lift}} =
\begin{bmatrix}\bar{{\mathcal{G}}}(x_2)&\cdots&\bar{{\mathcal{G}}}(x_{T+1})\end{bmatrix}.
\end{gather*}
Then the matrices $\mathcal{A}$ and $\mathcal{B}$ can be obtained by solving a least square  optimization problem:
\begin{gather}\label{stdform1}
    \min_{\mathcal{A},\mathcal{B}} {\lvert\lvert \mathbf{X}^{+}_{\text{lift}} - {\mathcal{A}}\mathbf{X}_{\text{lift}} - {\mathcal{B}}\mathbf{U}\rvert\rvert}_{\text{F}},
\end{gather}
where ${\lvert\lvert \cdot \rvert\rvert}_{\text{F}}$ denotes the Frobenius norm of a matrix. The solution of (\ref{stdform1}) gives 
\begin{equation}\label{eq:AB}
[\mathcal{A}\;\; \mathcal{B}] = \mathbf{X}_{\text{lift}}^{+} [\mathbf{X}_{\text{lift}}\;\;\mathbf{U}]^{\dagger},
\end{equation} 
where $\dagger$ denotes the Moore–Penrose pseudo-inverse of a matrix. In standard EDMD formulation, the projection from the higher dimensional space to the original lower dimensional space is achieved by $\mathcal{C}\in \mathbb{R}^{n\times N}$, computed by solving: 
\begin{gather}\label{stdform2}
    \min_{\mathcal{C}} {\lvert\lvert\mathbf{X} - \mathcal{C}\mathbf{X}_{\text{lift}}\rvert\rvert}_{\text{F}},
\end{gather}
which yields $\mathcal{C} = \mathbf{X} \mathbf{X}_{\text{lift}}^\dagger$. 

The computation of (\ref{eq:AB}) is sometimes prohibitive for large data sets $(T>>N)$, in which case it is beneficial to instead solve a ``normal form" equation obtained by multiplying 
$V := \begin{bmatrix}
    \mathbf{X}_{\text{lift}} \\ \mathbf{U}
    \end{bmatrix}\begin{bmatrix}
    \mathbf{X}_{\text{lift}} \\ \mathbf{U}
    \end{bmatrix}^\top$ on both sides of (\ref{eq:AB}) to obtain
$[\mathcal{A}\;\; \mathcal{B}] V=W$, where
   $ W := \mathbf{X}^{+}_{\text{lift}}\begin{bmatrix}
    \mathbf{X}_{\text{lift}} \\ \mathbf{U}
    \end{bmatrix}^\top.$ Then $[\mathcal{A}\;\; \mathcal{B}]=WV^\dagger$, in which
the dimension of $W$ and $V$ are $N\times(N+m)$, and $(N+m)\times(N+m)$, respectively, and are independent of the number of data samples $T$. Analogously, we can solve the normal form associated with $\mathcal{C} = \mathbf{X} \mathbf{X}_{\text{lift}}^\dagger$ to get $\mathcal{C} = \widehat{W} \widehat{V}^\dagger$, with $\widehat{W} = \mathbf{X}\mathbf{X}_{\text{lift}}^{\top}$, and $\widehat{V} = \mathbf{X}_{\text{lift}}\mathbf{X}_{\text{lift}}^{\top}$ having dimensions $n\times N$ and $N\times N$, respectively, that are again indendent of $T$.

\subsubsection{Proposed Method} Instead of relying on any predefined dictionary of basis function choices, here we explore the power of AI in data-driven learning of the basis functions, represented as deep neural networks (DNNs), for example, in \cite{robotics1}. Viewing the linear embedding to be a special case of a DNN, our approach learns not only the lifting functions (mapping to a higher dimension) but also the linear embedding (in the higher dimension), as well as the projections (mapping to the original lower dimension), all three in one go.

\begin{figure}[htbp!]
  \centering
    \includegraphics[width=0.48\textwidth]{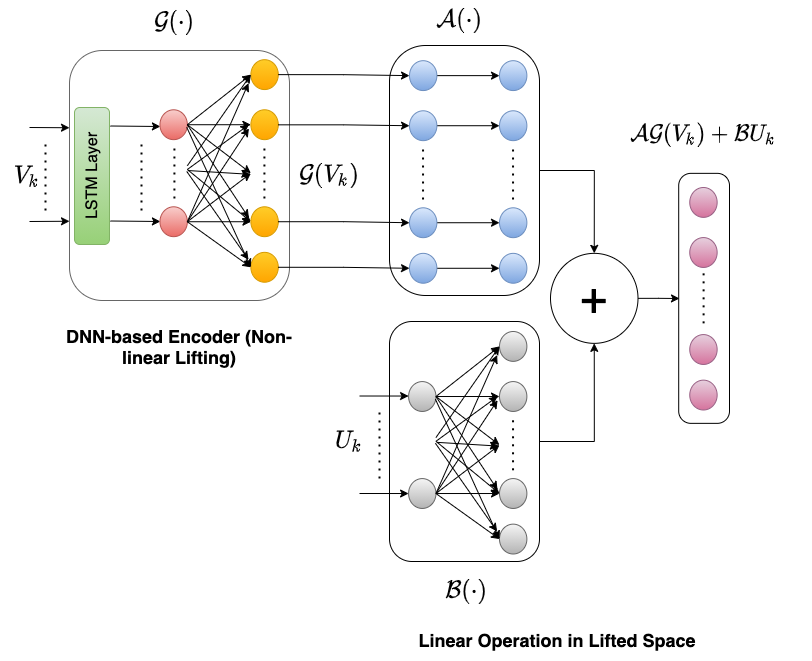}
  \caption{KDNN architecture: Encoder part}
  \label{f1}
 \end{figure}

Following the Koopman operator theory that approximates (\ref{geneqn2}) as (\ref{koopcon}), we can write the voltage dynamics (\ref{voleqn}) in the lifted linear space as:  
 \begin{equation}\label{volkoopcon}
\mathcal{G}(V_{k+1})\;\;=\;\;\mathcal{A}\; \mathcal{G}(V_{k}) \;+\; \mathcal{B}\; U_{k}.
\end{equation}
Our encoder-decoder KDNN architecture to realize (\ref{volkoopcon}) is presented in Fig.~\ref{f1},
where the matrix structure of $V_k$ of dimension $n \times H$  is provided in (\ref{volhist}), which can be flattened to form an vector input of length $n \times H$ at each control instant $k$. Since this input length can be long, increasing the complexity of the network to train, we extract the relevant temporal correlation by utilizing an LSTM (a type of recurrent neural network that can learn the long-term dependencies in a given time series). LSTMs capture the time dependencies of the data using three gates, namely the input gate, forget gate, and output gate, as detailed in \cite{lstm}. After the input LSTM layer, fully connected NN (FCNN) layers are used to train for the lifting, embedding linear dynamics, and projection (see Fig.~\ref{f1}). Note together the operation of the LSTM layer and the initial FCNN layer define the lifting function $\mathcal{G}(\cdot)$ as shown in Fig.~\ref{f1}. 

\begin{figure}[htbp!]
  \centering
    \includegraphics[width=0.40\textwidth]{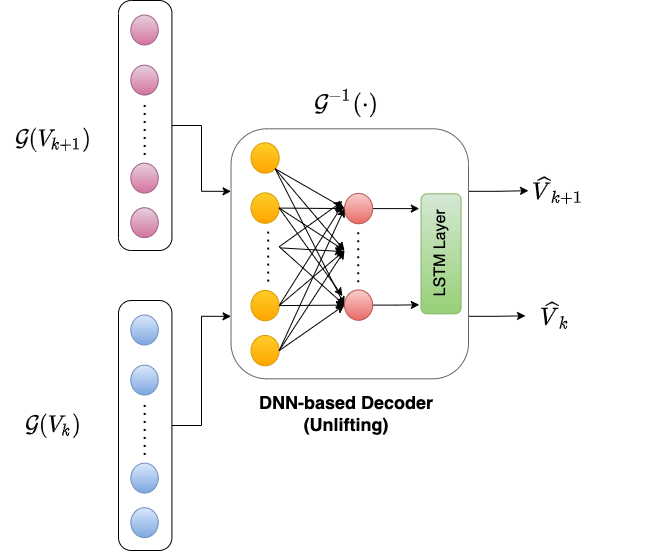}
  \caption{KDNN architecture: Decoder part}
  \label{f2}
 \end{figure}

Once lifted to a suitable higher dimension, the linear transformation of $\mathcal{G}(V_k)$ by the Koopman operator $\mathcal{A}(\cdot)$ is an instance of another FCNN (with identity activation function) to be trained as shown in Fig.~\ref{f1}. Similarly, the input $U_k$ is also processed through the Koopman operator $\mathcal{B}(\cdot)$, which is yet another instance of an FCNN with an identity activation function (see Fig.~\ref{f1}). The output of these two operation is added in the lifted space to obtain $\mathcal{A}\; \mathcal{G}(V_{k})+\mathcal{B}\; U_{k}]$, which equals $\mathcal{G}(V_{k+1})$ as per (\ref{volkoopcon}). Finally, to project to the original state-space, we utilize a decoder architecture shown in Fig.~\ref{f2}.
The decoder, denoted $\mathcal{G}^{-1}(\cdot)$, does the inverse operation of the encoder $\mathcal{G}(\cdot)$, and takes $\mathcal{G}(V_{k+1})=\mathcal{A}\; \mathcal{G}(V_{k}) + \mathcal{B}\;U_{k}$ as input and maps it to predicted $\widehat V_{k+1}$. Since we also have the signal $\mathcal{G}(V_{k})$, the same decoder map $\mathcal{G}^{-1}(\cdot)$ is also used to obtain the estimated version $\widehat V_{k}$ to compared against the ground truth $V_k$ and thereby ensure the accuracy of the decoder. Finally, the output layer is a cascade of an FCNN and an LSTM layer as the input layer but placed in reverse order. 

\begin{figure}[htbp!]
  \centering
    \includegraphics[width=0.48\textwidth]{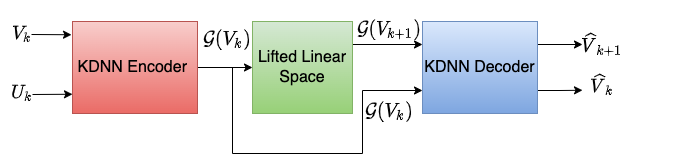}
  \caption{End-to-end KDNN architecture}
  \label{f3}
 \end{figure}

To summarize, the complete end-to-end architecture of the proposed KDNN is as shown in Fig.~\ref{f3}. 
The end-to-end KDNN is parameterized by $\theta$, and can be expressed as $\mathcal{F}(\cdot,\theta)$. The training input, output for the end-to-end KDNN are: $\mathcal{X} = \{V_k,U_k\},\mathcal{Y} = \{V_{k+1},V_k\}$, respectively, while the predicted output of the KDNN are $\widehat{\mathcal{Y}} = \{\widehat V_{k+1},\widehat V_k\}= \mathcal{F}(\mathcal{X},\theta)$.  We used mini-batch stochastic gradient descent to train $\mathcal{F}(\cdot,\theta)$. The loss function corresponds to the reconstruction error of both $V_{k+1}$ and $V_k$ is taken to be: 
\[\mathcal{L} (\theta) = \frac{1}{L}\sum_{i=0}^{L-1}{{\lvert\lvert \mathcal{Y}_i - \widehat{\mathcal{Y}_i} \rvert\rvert}_2},\] 
where $L$ denotes the batch size.

\subsection{Formulation of MPC problem}
Here our objective of the control design is to minimize the post-disturbance voltage deviations with respect to a user-defined reference value $V_{\text{ref}}$. For this, we employ MPC, which by design computes, at each control instant, an optimal sequence of control inputs for the remaining control horizon by optimizing a predicted future behavior of the underlying system, implements the first control action of the computed sequence, and then repeats the same process with the new measurements at the next control instant \cite{jin2009model}. This iterative control computation, employing new state measurements, helps to correct the effects of modeling error as compensated by the new measurement taken. Using the trained KDNN architecture, we can extract the functions $\mathcal{G(\cdot)}, \mathcal{A(\cdot)}$, and $\mathcal{B(\cdot)}$ that are then utilized in the computation of the MPC in the lifted linear space as formulated next. 


The voltages $V_k$, known from the measurements at control instant $k$ can be lifted to $\mathcal{G}(V_k)$. For the optimization formulation, we also lift the reference voltage $V_{\text{ref}}$ to obtain $\mathcal{G}(V_{\text{ref}})$. Then at any control instant $k$, the model-predictive optimization problem that needs to be solved is an instance of LQR (linear quadratic regulator):
\begin{subequations}\label{koopmpc}
\begin{multline}
\hspace*{-.2in}\min_{{U_{k}, \cdots, U_{k+N_k-1}}}
\!\!\!\!\sum_{i=0}^{N_k-1}\!\!\! \Big[(\mathcal G(V_{k+i+1})-\mathcal G(V_{\text{ref}}))^T\! Q(\mathcal G(V_{k+i+1})-\mathcal G(V_{\text{ref}}))\\[-6pt]
+U_{k+i}^T\!R U_{k+i}\Big]\nonumber
 \end{multline}
 \vspace*{-.2in}
\begin{align}
  \mbox{s.t.}\quad\mathcal G(V_{k+i+1}) =\mathcal{A}\;\mathcal G(V_{k+i}) + \mathcal{B}\; U_{k+i}, \;\;\;\forall i\in[0,N_k-1],\nonumber \\
    U_{\text{min}} \leq U_{k+i} \leq U_{\text{max}}, \;\;\;\forall i\in[0,N_k-1],\nonumber
\end{align}
\end{subequations}
\normalsize
where $N_k$ is the number of remaining control instants at the instant $k$, $Q \in \mathbb{R}^{N\times N}$ and $R \in \mathbb{R}^{m\times m}$ are the user specified weight matrices (with $N$ being the dimension of lifted space, and $m$ being the dimension of the control space), and $U_{\text{max}}$ and $U_{\text{min}}$ are the  lower/upper control bounds.

\section{Simulation Results}\label{3}
Our proposed methodology is validated through application to the IEEE 39-bus system to generate a closed loop MPC-based voltage stabilization policy following a 3-phase fault, to keep the voltage trajectories close to a reference value $V_{\text{ref}}=1.00$ p.u. under minimal control effort. The critical part of this design is training the KDNN architecture and learning the linear embedding for the voltage dynamics. 

\begin{figure}[t]
  \centering
    \includegraphics[width=0.45\textwidth]{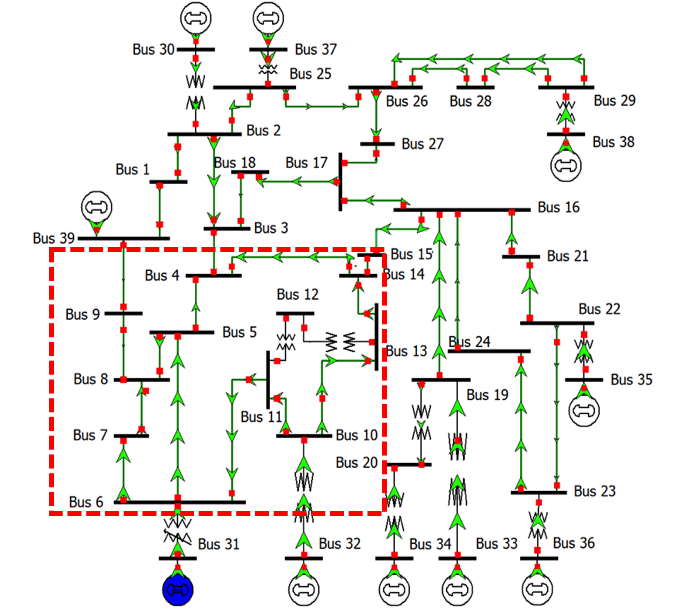}
  \caption{IEEE 39-bus System}
  \label{f4}
 \end{figure}

To formulate the control problem, we consider a fault at bus 15, which gets cleared by tripping a line between bus 15 and 16 within six cycles of system operation. Following the bus fault, voltages in the vicinity of the fault bus drop below the desired level almost immediately and need stabilization to avoid any potential collapse. Since the tripping of the line between buses 15 and 16 creates a disconnect between the region of Fig. \ref{f4} marked by the dashed red lines and the rest of the system, the voltages of the buses in the marked region are most affected, namely the voltages in the set: $\mathcal{S} = \{4, 5, 6, 7, 8, 9, 10, 11, 12, 13, 14, 15\}$. It then suffices to provide reactive compensation in the marked region, implying that the number of buses of interest is $n = \lvert\mathcal{S}\rvert = 12$. For the control of this region, we consider reactive control inputs at buses 4, 5, 7, 8, and 15 that can provide fast reactive support of up to 0.25 p.u.in each control step. To stabilize the voltage within 15 sec, the control horizon is set as 15 secs, divided into 5 segments of 3 secs each, i.e., $T_c=3$, and there are a total of $15/3=5$ control instants. 
The sampling duration is smaller by a factor of 4, i.e., $T_s=T_c/4=3/4$ sec, which implies the length of time-series between two control samples is $H=T_c/T_s=4$.

For training the proposed KDNN, we created a large pool of training data by simulating the system voltage trajectory under various load conditions within $\pm 10\%$ of the nominal load and also under different possible length-5 sequences of controls for each load condition, where the controls at each instant were chosen randomly in the range $[0,\;0.25$~p.u.]. In total, we created 2500 random load conditions, 3 different controls at each of the 5 control steps, collecting a total of $2500 \times 3 \times 5 = 37500$ data samples in the form of $\{(V_k,U_k,V_{k+1})\}$ triples. We divided the data into $70:30$ to create the training and test data sets. 

As shown in Fig.~\ref{f1}, the lifting function $\mathcal{G}(\cdot)$ is built using an LSTM layer followed by an FCNN layer, where the number of neurons of the FCNN layer decides the dimension $N$ of the lifted space. Given that $n=12$, we tested  $N = 64 > 5n$. The input and output layers of the KDNN employ non-linearity in the form of the $\tanh(\cdot)$ activation function for the LSTM as well as the FCNN layers, whereas the matrices $\mathcal{A}$ and $\mathcal{B}$  are represented as single layer FCNNs with identity as the activation function. The optimizer chosen for the training is ADAM, with gradient momentum $\beta_1=0.9/0.95$ and RMS momentum $\beta_2=0.999/0.95$. The loss function, batch size, learning rate, and performance metric are: mean squared error (MSE) loss, 32, $10^{-3}$, and mean absolute error (MAE), respectively. To facilitate fast and efficient learning and avoid over-fitting, we adopted the standard practice of pre-processing the input and output data of the KDNN: Firstly, $V_{\text{ref}} = 1.00$ was subtracted from the voltage values $V_k$ and $V_{k+1}$ of the stored data. Next, this modified data was normalized in the range $[0,1]$ with respect to the modified data set's minimum and maximum values. The control values $U_k$ in the range of $[0,0.25]$ were normalized between $[-1,1]$. 

\begin{figure}[htbp!]
  \centering
    \includegraphics[width=0.45\textwidth]{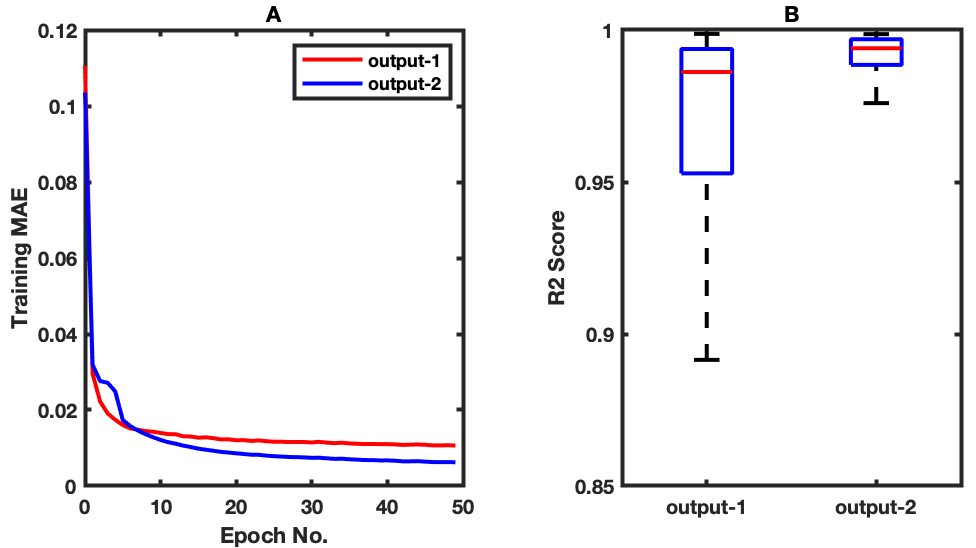}
  \caption{KDNN Training and Testing Performance for $N=64$}
  \label{f5}
 \end{figure}

The training and testing performances of reconstruction of $V_{k+1}$ (output-1) and $V_{k}$ (output-2) are shown in Fig.~\ref{f5} in terms of MAE, and in Fig.~\ref{f4}.B in terms of the coefficient of determination or $R^2\in[0,1]$ score. (Recall an $R^2$ value of 1 indicates an exact fit.) 
It should be noted that both training and testing performances are satisfactory; hence we utilized the trained KDNN to extract the functions $\mathcal{G(\cdot)}, \mathcal{A(\cdot)}$, and $\mathcal{B(\cdot)}$ for MPC computation in the lifted linear space. 

Considering the objective of minimizing the voltage deviations from a reference, we chose $Q=I_{N\times N}$ and $R=0$. For validating the robustness of the proposed control design, we considered 5 different load levels 90\%, 95\%, 100\%, 105\%, and 110\% of the nominal load. In addition to the fault at Bus-15, we considered faults at Bus-4, Bus-7, and Bus-8. These faults were cleared by tripping the transmission lines between Bus-3 and Bus-4, Bus-6 and Bus-7, and Bus-7 and Bus-8, respectively. 

\begin{figure}[htbp!]
\centering
\includegraphics[width=0.48\textwidth]{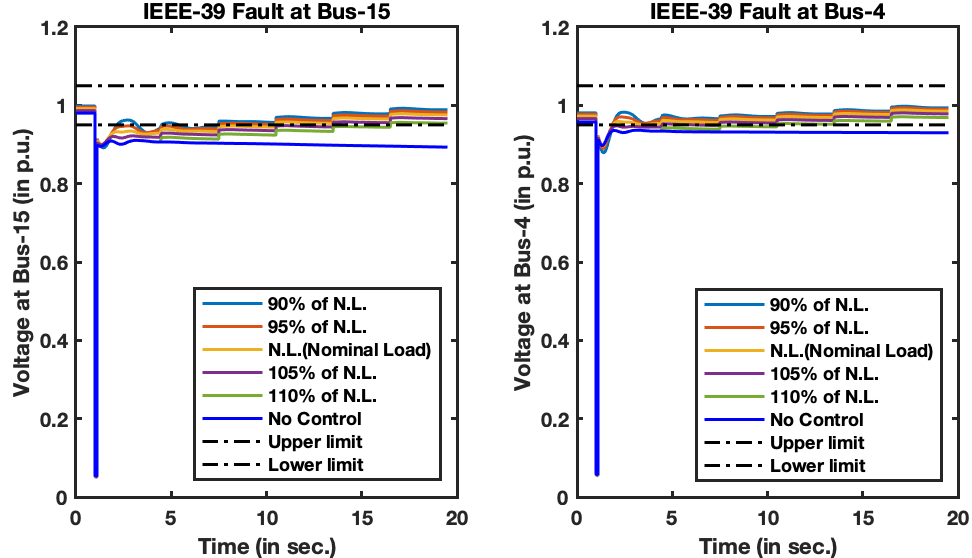}
\caption{Voltage plots for faults at Bus-15 and Bus-4}
\label{f6}
 \end{figure}

\begin{figure}[htbp!]
\centering
\includegraphics[width=0.48\textwidth]{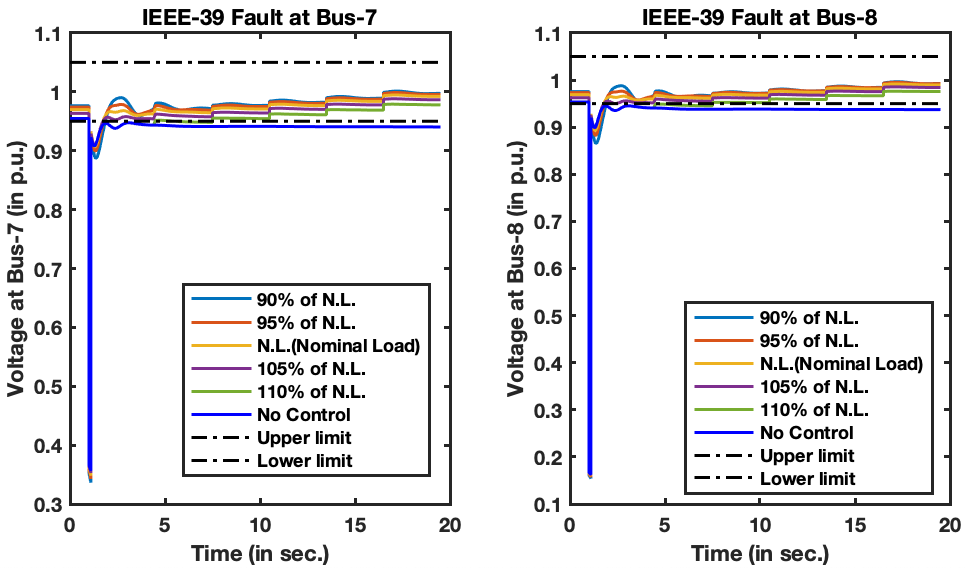}
\caption{Voltage plots for faults at Bus-7 and Bus-8}
\label{f7}
\end{figure}

The MPC computations were done in the linear embedded state-space, solving the LQR of Section II.C. Next, the computed controls were applied to the original nonlinear system. The voltage profiles for each of the above cases are shown in Fig.~\ref{f6} and Fig.~\ref{f7}, validating that the proposed scheme successfully achieved the desired voltage performance under different operating conditions, thereby confirming the effectiveness and robustness of the proposed methodology of designing controls using the KDNN-based lifted linear embedding of the nonlinear dynamics. 

\begin{figure}[htbp!]
\centering
\includegraphics[width=0.48\textwidth]{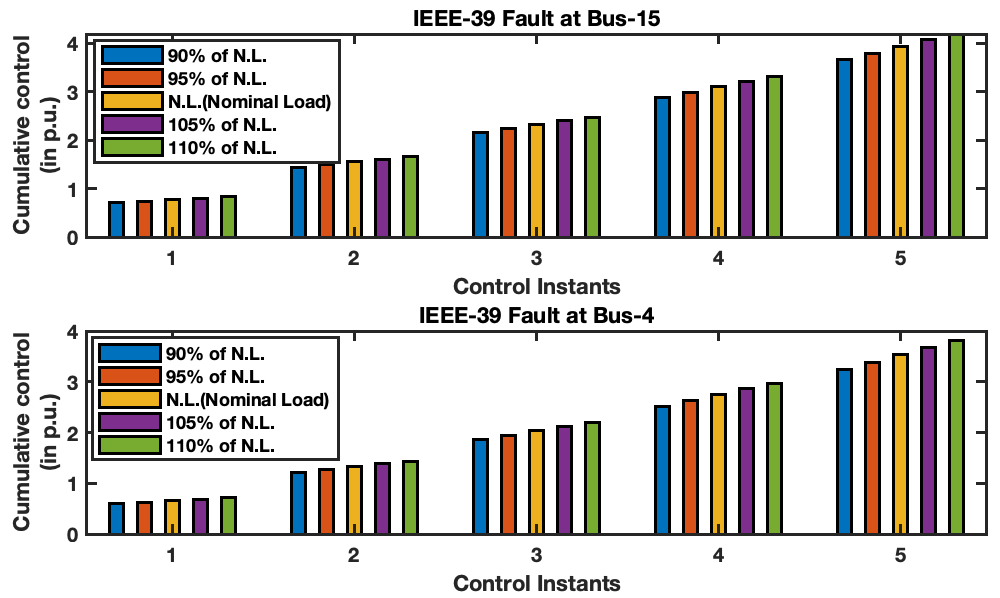}
\caption{Cumulative control plots for faults at Bus-15 and Bus-4}
\label{f8}
 \end{figure}

\begin{figure}[htbp!]
\centering
\includegraphics[width=0.48\textwidth]{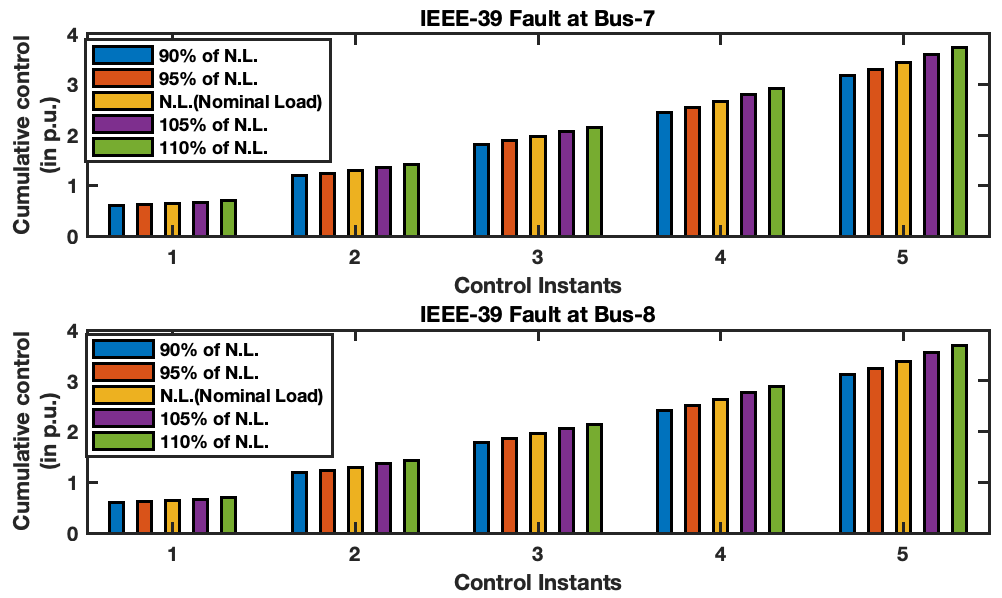}
\caption{Cumulative control plots for faults at Bus-7 and Bus-8}
\label{f9}
\end{figure}

We also computed the control actions at each control instants and plotted the accumulated control actions for the 5 different load cases in Fig.~\ref{f8} and Fig.~\ref{f9}. The trend suggests that with the load increase, the amount of VAR compensation increased, and this trend is expected.

\section{Comparison with Standard Approaches}
\subsection{Prediction performance: Standard Koopman vs. KDNN}
As mentioned in \cite{robotics1}, an early work in NN-based Koopman design, radial basis function, polynomials, and kernel functions are mostly common basis functions in Koopman-operator-based control design. But the choice of an appropriate basis function is an open problem, and this is our primary motivation for utilizing NNs to use data to learn the appropriate basis functions. Here, we demonstrate the benefit of the proposed KDNN-based design compared to the standard EDMD method mentioned in Section-II.B for approximating nonlinear implicit voltage dynamics. 
\begin{figure}[htbp!]
  \centering
    \includegraphics[width=0.45\textwidth]{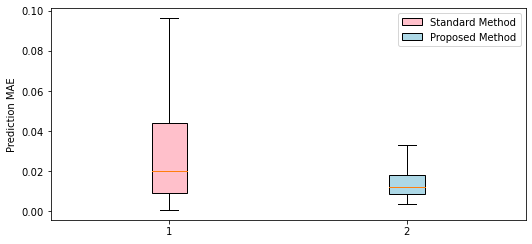}
  \caption{Comparison of MAE for standard method and proposed method}
  \label{comp_std}
 \end{figure}

First, note that for the polynomial basis, the choice of candidate basis set increases double-exponentially with the dimension of original nonlinear dynamics, $n\times H$, per the problem formulation. This makes creating an optimal choice of dictionary non-scalable and hence prohibitive. 
 
So for comparison, we picked radial basis functions with a dictionary size of 2000. The centers of the radial basis functions were determined by K-means clustering over training data, while for the spread parameter, we selected $\sigma = 0.05$. We then followed (\ref{stdform1})-(\ref{stdform2}) to find $\mathcal{A}$, $\mathcal{B}$, and $\mathcal{C}$ over the same training data utilized to train the KDNN. We then used the computed matrices to find the predicted values of the voltage dynamics over the same testing data used for the KDNN testing. The mean absolute errors of both predictions are plotted in Figure~\ref{comp_std}, which clearly shows the superiority of our KDNN-based learned basis functions. 

\subsection{Computation time: State-of-art MPC vs KDNN MPC}
The comparison of online control computation time for trajectory-sensitivity based state-of-art MPC \cite{jin2009model} and the proposed KDNN-based MPC is shown in Table \ref{tab:time}. It demonstrates an impressive 35-fold speed-up of the proposed scheme over the state-of-art MPC implementation. Also, the proposed method takes 0.2~s to compute a control at each online decision instant, comparable to the one used in practice, making MPC real-time and practical for power systems. It is important to note that even the traditional controllers, e.g., UVLS relaying scheme, generally needs $\sim$0.5~s to decide a control action \cite{dong2017emergency}.
\begin{table}[htbp!]
\caption{Comparison of computation time}\label{tab:time} 
\centering
\tabcolsep=0.5 cm
\begin{tabular}{| c | c |}
\hline
\textbf{Method} & {\textbf{Average Time}} \\
\hline
MPC in \cite{jin2009model} & 7.00 sec/step  \\ \hline
Proposed Method & 0.20 sec/step\\ \hline
\end{tabular}
\end{table}
For our implementation and computation, we used intel(R) Core(TM) i7-4790 CPU @ 3.60GHz processor with 16 GB RAM.

\section{Conclusions}\label{sec:conclusion}
The paper proposed and implemented a Koopman-inspired encoder-decoder framework for the data-driven linear embedding of the dynamics of distributively controlled networked systems, paving the way for designing a control strategy in the lifted linear state-space, making the MPC design scalable and real-time for a first time for power networks. We combined the concept of Koopman operator theory for lifting nonlinear dynamics into higher dimensional linear dynamics with the power of deep learning to learn in one go the lifting and projection functions of the encoder-decoder as well as the high dimensional linear embedding. This data-driven approach auto-learns the basis/projection functions removing the burden of selecting those arbitrarily, traditionally taken to be polynomials or radial bases. The test results applied to the IEEE 39-bus system validated the performance of the newly proposed scheme in terms of efficacy, robustness against load variations, and fault conditions. We also validated the superiority of our approach compared to the standard EDMD approaches for Koopman-embedding, which employ pre-defined basis functions. The proposed promising technique of unraveling the implicit nonlinear dynamics, combining Koopman theory and deep learning methods, is general, opening up a new direction of distributed control design for complex nonlinear networked systems.


%



%

\bibliographystyle{IEEEtran}
\bibliography{IEEEabrv,Bibliography}

%




\end{document}